\documentclass[preprint]{aastex62}

\newcommand{\speed}[1]{#1 km~s${}^{-1}$}
\newcommand{\accel}[1]{#1 km~s${}^{-2}$}

\newcommand{\nfig}[1]{Figure~\ref{#1}}

\usepackage{amsmath}

\received{... ..., 2019}
\revised{... ..., 2019}
\accepted{... ..., 2019}
\submitjournal{ApJ}

\shorttitle{Round-Trip Slipping Magnetic Reconnection Observed in a Fan-Spine Jet}
\shortauthors{Shen et al.}

\begin{document}

\title{Round-trip Slipping Motion of the Circular Flare Ribbon Evidenced in a Fan-Spine Jet}

\correspondingauthor{Yuandeng Shen}
\email{ydshen@ynao.ac.cn}

\author[0000-0001-9493-4418]{Yuandeng Shen}
\affiliation{Yunnan Observatories, Chinese Academy of Sciences,  Kunming, 650216, China}
\author{Zhining Qu}
\affiliation{School of Automation and Information Engineering, Sichuan University of Science \& Engineering, Zigong 643000, China}
\author{Chengrui Zhou}
\affiliation{Yunnan Observatories, Chinese Academy of Sciences,  Kunming, 650216, China}
\author{Yadan Duan}
\affiliation{Yunnan Normal University, Department of Physics, Kunming 650500, China}
\author{Zehao Tang}
\affiliation{Yunnan Observatories, Chinese Academy of Sciences,  Kunming, 650216, China}
\author{Ding Yuan}
\affiliation{Institute of Space Science and Applied Technology, Harbin Institute of Technology, Shenzhen, 518055, China}

\begin{abstract}
A solar jet on 2014 July 31, which was accompanied by a {\em GOES} C1.3 flare and a mini-filament eruption at the jet base, was studied by using observations taken by the New Vacuum Solar Telescope and the  {\em Solar Dynamic Observatory}. Magnetic field extrapolation revealed that the jet was confined in a fan-spine magnetic system that hosts a null point at the height of about 9 Mm from the solar surface. An inner flare ribbon surrounded by an outer circular ribbon and a remote ribbon were observed to be associated with the eruption, in which the inner and remote ribbons respectively located at the footprints of the inner and outer spines, while the circular one manifested the footprint of the fan structure. It is interesting that the circular ribbon's west part showed an interesting round-trip slipping motion, while the inner ribbon and the circular ribbon's east part displayed a northward slipping motion. Our analysis results indicate that the slipping motions of the inner and the circular flare ribbons reflected the slipping magnetic reconnection process in the fan quasi-separatrix layer, while the remote ribbon was associated with the magnetic reconnection at the null point. In addition, the filament eruption was probably triggered by the magnetic cancellation around its south end, which further drove the slipping reconnection in the fan quasi-separatrix layer and the reconnection at the null point.
\end{abstract}

\keywords{Sun: activity --- Sun: flares --- Sun: filaments, prominences --- Sun: chromosphere --- Sun: magnetic fields} 

\section{Introduction}
Jet or jet-like phenomena are widely observed in the solar atmosphere; they are thought to be important for understanding the enigmatic problems of coronal heating and the acceleration of solar winds \citep{2016SSRv..201....1R}. Based on morphology, solar jets are divided into collimated anemone jets and two-sided-loop jets, in which the former is produced by the magnetic reconnection between emerging bipoles and the ambient open magnetic fields \citep{1995Natur.375...42Y,2011ApJ...735L..43S}, while the latter is by the reconnection between emerging bipoles and the overlying horizontal fields \citep{1995Natur.375...42Y,2017ApJ...845...94T,2018ApJ...869...39M,2019ApJ...871..220S}. In addition, solar jets are also important to trigger other kind of large-scale solar eruptions such as coronal waves \citep[e.g.,][]{2018ApJ...861..105S,2018MNRAS.480L..63S,2018ApJ...860L...8S}, filament oscillations \citep[e.g.,][]{2014ApJ...785...79L,2017ApJ...851...47Z,2019ApJ...872..109A}, and coronal mass ejections \citep[e.g.,][]{2008ApJ...677..699J,2012ApJ...745..164S,2019ApJ...881..132D}. Recent observations have revealed that the majority of collimated and two-sided-loop jets are all driven by erupting mini-filaments \citep{2012ApJ...745..164S,2017ApJ...851...67S,2019ApJ...883..104S,2015Natur.523..437S,2019ApJ...871..220S}, and new models have been proposed to interpret these new observational features \citep{2010ApJ...720..757M,2012ApJ...745..164S,2019ApJ...883..104S}. 

Coronal jets are frequently occurring in fan-spine magnetic topologies associated with parasitic polarities \citep[e.g.,][]{2009ApJ...700..559M,2011ApJ...728..103L,2013ApJ...778..139S,2015ApJ...812...50J,2015ApJ...806..171Y,2015ApJ...809...45Z,2016ApJ...830...60H,2017ApJ...835...35H,2016NatCo...711522J,2017ApJ...836..235L,2017ApJ...842L..20L,2019ApJ...872...87L,2017A&A...605A..49C,2017ApJ...837..173R,2018ApJ...860L..25Y,2019ApJ...871....4H}.  A fan-spine topology is a complicated magnetic system composed of a coronal null point, a dome-like fan portrays the closed separatrix surface, and the inner and outer spine field lines belong to different connectivity domains \citep{1990ApJ...350..672L,2009ApJ...704..485T}. Flare ribbons are elongated bright emission features attributed to low-altitude impact of particle beams accelerated through magnetic reconnection, whose locations reflect the footprints of separatrices or quasi-separatrix layers \citep[QSLs,][]{1997A&A...325..305D,1997SoPh..174..229M}. Observationally, within the framework of fan-spine topologies, one can expect three flare ribbons in an eruption, namely, an inner brightening surrounded by a circular ribbon relevant to the inner spine and the dome-like fan structures, and a remote ribbon associated with the outer spine. 

The kinematics of flare ribbons can supply information for diagnosing the key physical process in solar eruptions \citep{1996RSPTA.354.2951P}. Since flare ribbons in chromosphere manifest the footprint of QSLs that too tend to harbor high electric current density, the locations of flare ribbons are useful evidence for investigating the topology structures of coronal magnetic fields. Flare ribbons often show three types of motions: converging, separation, and slipping. Converging and separation motions of conjugated flare two ribbons are perpendicular to the magnetic neutral line, and they are interpreted to be manifestations of the downward and upward motions of the reconnection sites \citep[e.g.,][]{2006ApJ...636L.173J}. Slipping motions are along flare ribbons, which manifest the slipping magnetic reconnection process within QSLs \citep{2006SoPh..238..347A,2014ApJ...784..144D,2016ApJ...823...41D,2016ApJ...823..136Z,2016ApJ...830..152L,2018ApJ...859..122L,2019ApJ...881...68L}. For fan-spine magnetic topologies, the flare ribbons at different locations are caused by energetic particle beams along different magnetic loops as mention above \citep{2012A&A...547A..52R}. So far, only one study reported the round-trip slipping motion along the inner flare ribbon \citep{2012ApJ...760..101W}. The slipping motion of  circular flare ribbons towards a certain direction have been observed in a few cases \citep[e.g.,][]{2009ApJ...700..559M,2013ApJ...774...60L,2015PASJ...67...78J,2017ApJ...837..173R,2017ApJ...850..167L,2018ApJ...859..122L}, but round-trip slipping motion of circular flare ribbon has not been reported yet.

Here, we report the first fan-spine jet in which the circular flare ribbon exhibited round-trip slipping motion, using high-resolution H$\alpha$ images taken by the New Vacuum Solar Telescope \citep[NVST;][]{2014RAA....14..705L} and the {\sl Solar Dynamic Observatory} \citep[{\sl SDO};][]{2012SoPh..275....3P}. The NVST H$\alpha$ images have a cadence of 6 seconds with a pixel resolution of {0\arcsec.16}. The extreme ultraviolet (EUV) images taken by the {\em SDO}'s Imager and Atmospheric Imaging Assembly \citep[AIA;][]{2012SoPh..275...17L}, and the line-of-sight (LOS) and vector magnetograms taken by the {\em SDO}'s Helioseismic and Magnetic Imager \citep[HMI;][]{2014SoPh..289.3483H} are also used in this paper. The pixel size and cadence of the AIA EUV images (HMI LOS magnetograms) are {0\arcsec.6} ({0\arcsec.5}) and 12 (45) seconds, respectively. 

\section{Results}
\subsection{Pre-eruption Magnetic Condition}
On 2014 July 31, a solar jet, which was accompanied by a {\em GOES} C1.3 flare and the eruption of a mini-filament within the eruption source region, occurred at the west periphery of active region AR12127. The flare's start, peak, and end times were 02:04, 02:08, and 02:10 UT, respectively.

\begin{figure}
\epsscale{1}
\figurenum{1}
\plotone{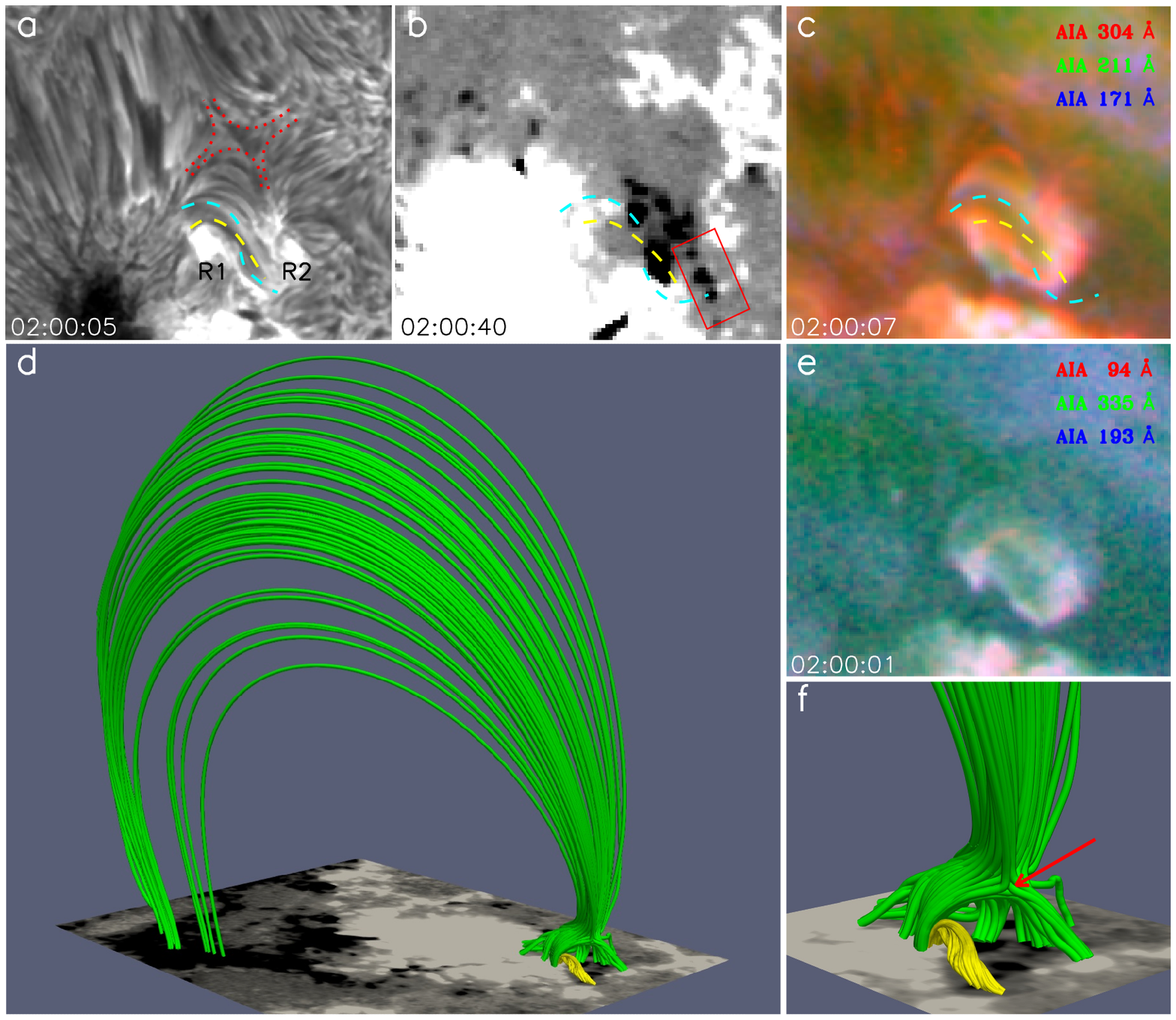}
\caption{Panel (a) is a NVST H$\alpha$ image, in which the dotted curves outline the X-shaped structure, while the dashed curves indicate the twisted mini-filament. 'R1' and 'R2' mark the two bright ribbons on the both sides of the filament. Panel (b) is an HMI LOS magnetogram in which the white (black) patches are positive (negative) polarities. Panels (c) and (e) are composite AIA images; they are made from AIA's low temperature channels of 304 \AA\, 211 \AA\ , and 171 \AA\ and high temperature channels of 94 \AA\, 193 \AA\, and 335 \AA\, respectively. The yellow and cyan curves in panel (a) are also overlaid in panels (b) and (c). Panel (d) shows some representative field lines of the fan-spine structure (green) and the mini-filament (yellow), and the bottom is an HMI LOS magnetogram. Panel (f) is the  closeup view of the fan structure and the mini-filament, in which the red arrow points to the  null point structure. The field-of-view (FOV) for panels (a)--(c) and (e) is $60\arcsec \times 50\arcsec$.
\label{fig1}}
\end{figure}

An overview of the pre-eruption magnetic condition is shown in \nfig{fig1}. In the H$\alpha$ image (\nfig{fig1}(a)), a twisted mini-filament outlined by the dashed curves and an X-shaped structure marked by the dotted curves can be identified. The filament lied on the polarity inversion line, whose north (south) end rooted in positive (negative) magnetic field region (see \nfig{fig1}(b)). The filament can also be observed in the composite low and high temperature images (see \nfig{fig1}(c) and (e)). To make a composite image, three near-simultaneous AIA images at different wavelengths are put into the red, green, and blue channels of a true color image. Here, the strong response to logarithmic temperature of the high and low temperature composite images are in the ranges of 6.2--6.8 and 4.7--5.8, respectively.

To investigate the coronal magnetic condition around the jet eruption source region, the pre-eruption HMI vector magnetogram at 01:00:00 UT is used as the bottom boundary condition to extrapolate the three-dimensional coronal magnetic field using the nonlinear force-free magnetic field (NLFFF) code \citep{2000ApJ...540.1150W}. To perform the NLFFF extrapolation, the photospheric vector magnetogram was preprocessed to best suit the force-free condition \citep{2006SoPh..233..215W}, then the extrapolation is performed in the cubic box of $704 \times 480 \times 480$ uniformly spaced grid points with $\Delta x = \Delta y = \Delta z =$ 0\arcsec.5, by using the optimization method of \cite{2000ApJ...540.1150W} and \cite{2004SoPh..219...87W}. Based on the extrapolated coronal magnetic field, a fan-spine magnetic topology hosting a null point, and a mini-filament underneath the fan are identified (see \nfig{fig1}(d)). The closeup view around the eruption source region is shown in \nfig{fig1}(f), which clearly shows the locations of the null point (see the red arrow) and the twisted mini-filament (yellow lines) structures. By comparing the extrapolated coronal magnetic structure with the H$\alpha$ image, it is reasonable to believe that the X-shaped structure in the H$\alpha$ image should be the projection of the null point evidence in the extrapolated coronal magnetic field.

\subsection{Slipping Motions of Flare Ribbons}
\begin{figure}
\epsscale{1}
\figurenum{2}
\plotone{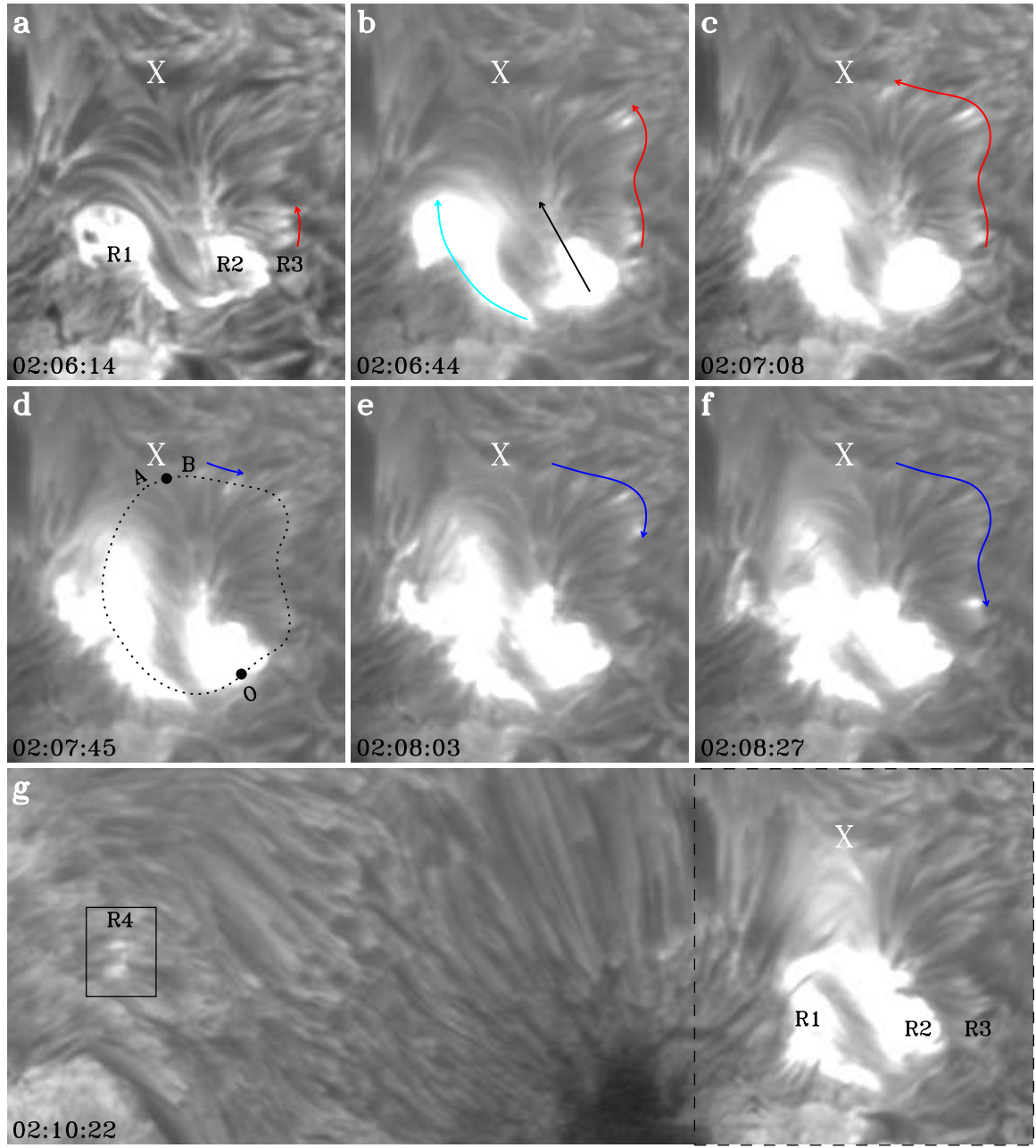}
\caption{The top two rows are NVST H$\alpha$ images. 'R1' and 'R2' in panel (a) mark the two bright ribbons on the both sides of the mini-filament, and their slipping directions are indicated respectively by the cyan and black arrows in panel (b). The red arrow in panels (a)--(c) show the northward slipping motion of the west part of the circular flare ribbon, and its southward slipping motion is indicated by the blue arrows in panels (d)--(f). Panel (g) shows a larger FOV that covers the remote ribbon (R4, see the black box). The symbol 'X' in each panel marks location of the X-shaped structure. The FOV for panels (a)--(f) is $34\arcsec \times 38\arcsec$, and that for panel (g) is $103\arcsec \times 38\arcsec$. The dashed box in panel (g) indicates the FOV of (a)--(f).
\label{fig2}}
\end{figure}

The flare was associated with three bright ribbons, i.e., an inner ribbon, an outer circular ribbon, and a remote ribbon. During the flare's rising phase, the inner and the circular ribbons showed obvious slipping motions. At about 02:04:07 UT, two bright patches appeared on the both sides of the mini-filament (see R1 and R2 in \nfig{fig2}(a)). About two minutes latter, the region on the west of R2 was also lighted up (see R3 in \nfig{fig2}(a)). Here, R1 and R3 made up the circular ribbon, and R2 is recognized as the inner ribbon. R1 and R2 showed obvious northward slipping motions (indicated by the cyan and black arrows in \nfig{fig2}(b)), and they finally ended at the location of the X-shaped structure.

It is interesting that R3 showed a round-trip slipping motion. It first slipped to the north (see the red arrows in \nfig{fig2}(a)--(c)). After the bright emission front reached up to the X-shaped structure, it started to slip to the south along its original trajectory (see the blue arrows in \nfig{fig2}(d)--(f)). At about 02:08:09 UT, the remote ribbon (R4) and the brightening at the X-shaped structure appeared (see the black box in \nfig{fig2}(g)), which was near simultaneously with the intruding of R1, R2, and R3 into the X-shaped structure. This may suggests that the R4 and the brightening at the X-shaped structure were caused by the null point reconnection (see the online animation associated with \nfig{fig4}).

\begin{figure}
\epsscale{0.9}
\figurenum{3}
\plotone{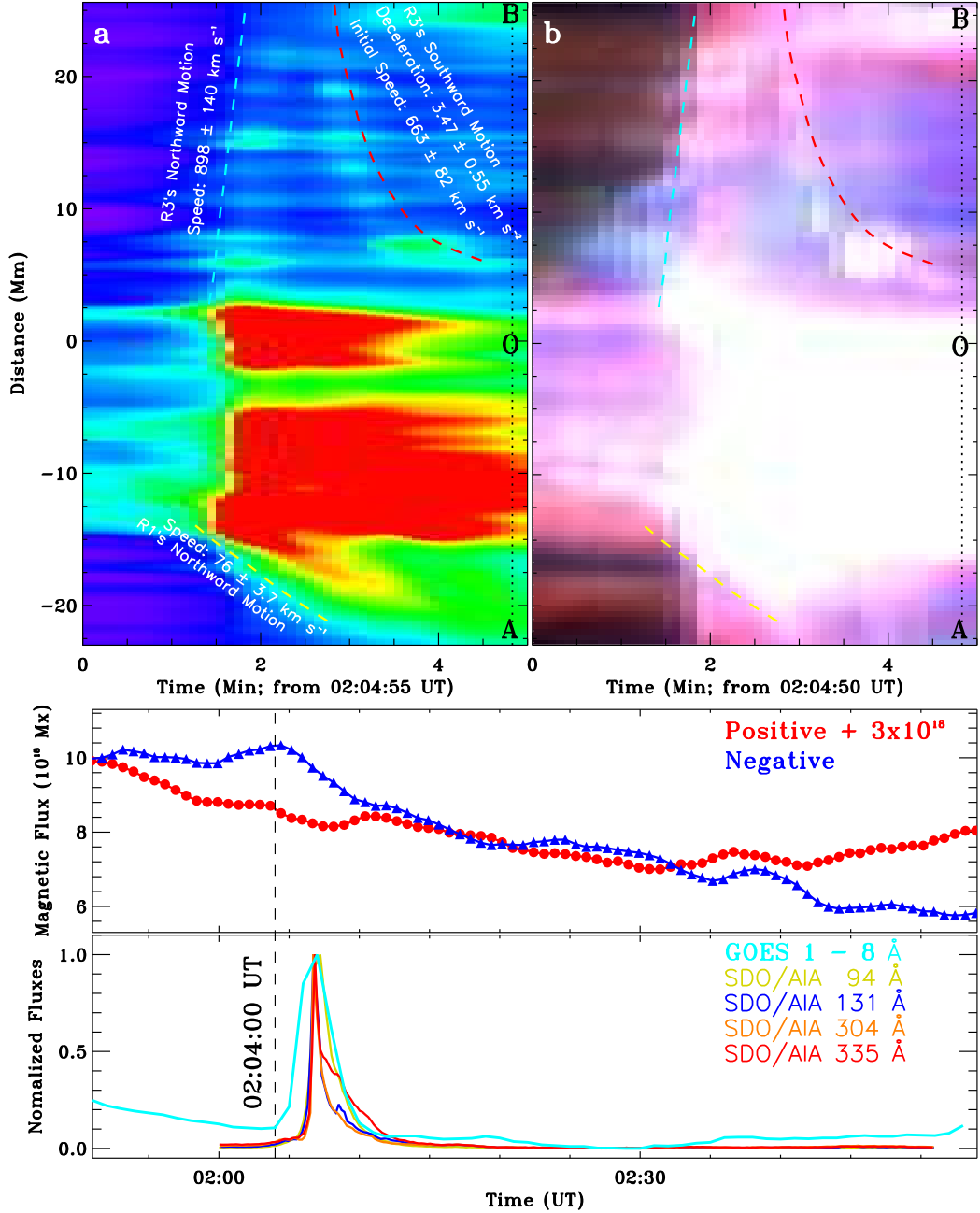}
\caption{Time-distance diagrams made along the black dotted curve in \nfig{fig2}(d). Panel (a) is made from the NVST H$\alpha$ images, and panel (b) is made from the composite images of AIA 94 \AA\, 131 \AA\, and 335 \AA\ images. Since the east and west parts of the circular ribbon moved in opposite direction, we set point O as the origin of coordinates in the time-distance diagrams (see panels (a), (b), and \nfig{fig2}(d)). The yellow dashed line in panel (a) indicates the northward slipping motion of the east part of the circular flare ribbon; the northward and southward slipping motions of the west part of the circular flare ribbon are indicated by the cyan dashed line and the red dashed curve, respectively. The curve and lines in panel (b) are the same with those in panel (a). Panel (c) shows the variations of the positive (red) and negative (blue) magnetic fluxes within the red box in \nfig{fig1}(b). Panel (d) shows the normalized {\em GOES} soft X-ray 1--8 \AA\ flux and the AIA lightcurves of AIA 94 \AA\ (yellow), 131 \AA\ (blue), 304 \AA\ (orange), and 335 \AA\ (red) within the black box in \nfig{fig4}(d).
\label{fig3}}
\end{figure}

The kinematics of the circular ribbon is analyzed using time-distance diagram made along the black dotted curve in \nfig{fig2}(d). \nfig{fig3}(a) shows the time-distance diagram made from H$\alpha$ images. Since R1 and R3 moved in opposite directions, we set point O as the zero point of the time-distance diagram's coordinate. Thus, OA and OB represent the motion directions of R1 and R3, respectively. By fitting the bright emission front with a linear function, it is obtained that the northward slipping speeds of R1 and R3 were about \speed{-76 $\pm$ 3.7 and 898 $\pm$ 140}, respectively. The southward slipping motion of R3 showed an obvious deceleration (see the red dashed curve in \nfig{fig3}(a)). Therefore, we fit it with a quadratic function and obtain that the deceleration was of about \accel{3.47 $\pm$ 0.55}. We further fit the data points during the initial stage (one third of the total data points) with a linear function and obtain that the initial speed of R3's southward motion was about \speed{663 $\pm$ 81}. \nfig{fig3}(b) shows the time-distance diagram made from the composite images made from the AIA 94 \AA\, 131 \AA\, and 335 \AA\ images, which showed similar moving characteristics as observed in the H$\alpha$ observations.

\subsection{Photospheric Magnetic Flux Variations and the Jet Eruption}
\begin{figure}
\epsscale{1}
\figurenum{4}
\plotone{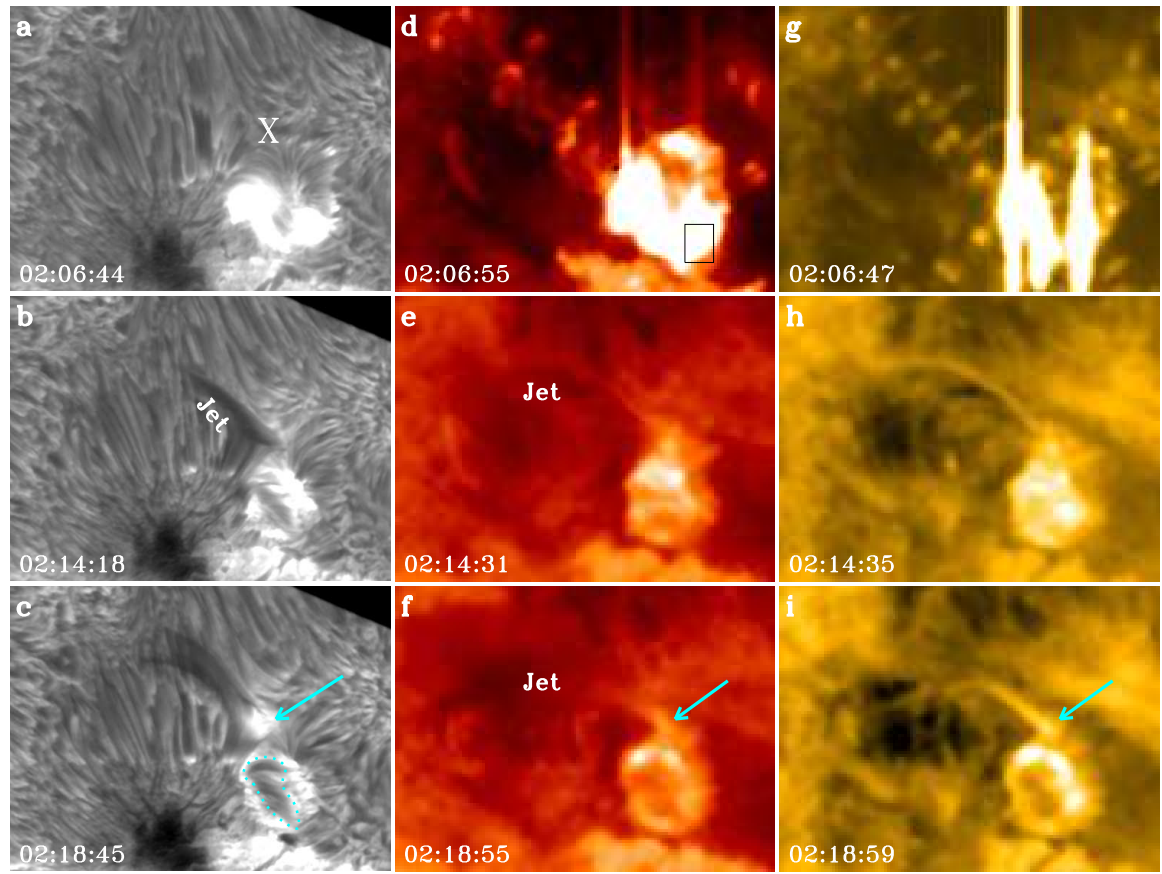}
\caption{The left column shows NVST H$\alpha$ images. The symbol 'X' marks the X-shaped structure. The middle and right columns show the AIA 304 and 171 \AA\ images, respectively. The brightening at the X-shaped structure is indicated by the cyan arrows in the bottom row. The FOV for each panel is $80\arcsec \times 60\arcsec$. An animation is available in the online journal, which runs from 02:00 to 02:36 UT.
\label{fig4}}
\end{figure}

In order to investigate the triggering mechanism of the eruption, the variations of the photospheric magnetic fluxes within the red box in \nfig{fig1}(b) (around the south end of the mini-filament) are analyzed, and the results are plotted in \nfig{fig3}(c). One can see that the positive and negative magnetic fluxes showed a sudden decrease at about 02:04:00 UT, which suggests the occurrence of flux cancellation between the opposite magnetic polarities. The {\em GOES} soft X-ray flux and AIA lightcurves within the black box in \nfig{fig4}(d) are plotted in \nfig{fig3}(d). It is found that the start time of the flux cancellation was coincided with the beginning of the flare. This result indicates that the initiation of the flare/eruption was probably triggered by the flux cancellation, through breaking of the mini-filament at its south end from the solar surface.

The jet eruption is displayed in \nfig{fig4} and the online animation. In the H$\alpha$ images, an inversed Y-shaped jet gradually appeared at about 02:10:10 UT after the brightening at the X-shaped structure, and it ejected along the outer spine as evidenced in the extrapolated magnetic field. At about 02:14:18, the jet body can well be identified in the H$\alpha$ and AIA images (see \nfig{fig4}(b), (e), and (h)). The brightening at the X-shaped structure is indicated by the cyan arrows in the bottom row of \nfig{fig4}. Here, the brightening at the X-shaped structure can be regarded as the magnetic reconnection signal at the coronal null point. The ejecting speeds of the jet  measured from the H$\alpha$ and AIA 304 \AA\ images were about \speed{100} and \speed{137}, respectively. By comparing the filament shapes before (\nfig{fig1}(a)) and after (\nfig{fig4}(c)) the jet, one can find that the mini-filament had became less twisted, which may imply the partial eruption of the filament. Since the flux cancellation broke the filament at the south end, the eruption of the filament should be from south to north and with its north end remained stable. 

\subsection{Squashing Factor and Magnetic Pressure}
\begin{figure}
\epsscale{1}
\figurenum{5}
\plotone{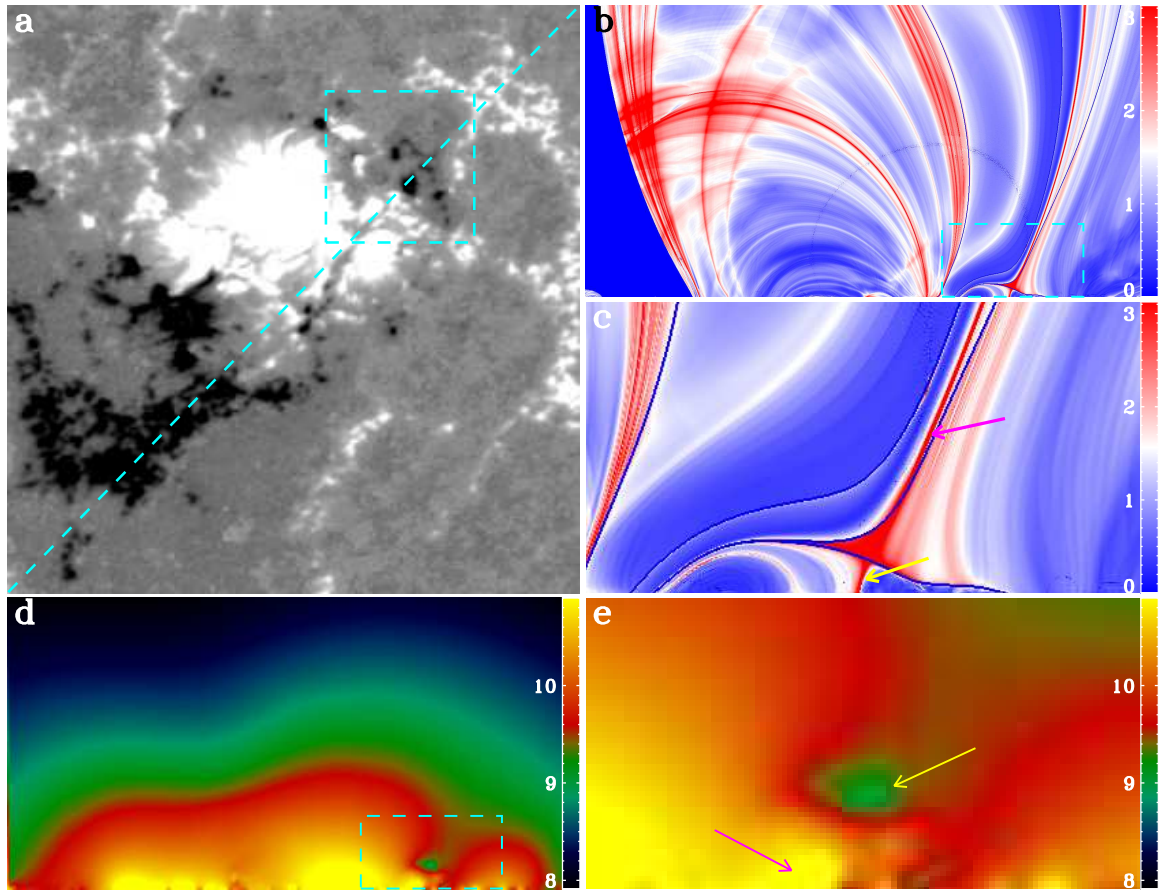}
\caption{Panel (a) is an HMI LOS magentogram in which the cyan dashed box indicates the eruption source region, while the dashed line indicates the location where we calculate the squashing factor Q (panel (b)) and magnetic pressure (panel (d)) in altitude. The closeup view of the box regions in panels (b) and (d) are plotted in panels (c) and (e), respectively. The pink and yellow arrows in panel (c) point to the outer and inner spines, respectively. The yellow and pink arrows in panel (e) point to the null point and the location of the mini-filament, respectively.
\label{fig5}}
\end{figure}

Using the code developed by \cite{2016ApJ...818..148L}, the squashing factor Q \citep{2002JGRA..107.1164T} and magnetic pressure maps are generated by using the extrapolated magnetic field. \nfig{fig5}(a) is an HMI magnetogram, in which the cyan dashed box indicates the eruption source region, while the dashed line indicates the location where we calculate the squashing factor Q (\nfig{fig5}(b)) and magnetic pressure (\nfig{fig5}(d)) in altitude. The closeup view of the box regions in panels (b) and (d) are plotted in panels (c) and (e), respectively. The Q map shows the fan-spine topology clearly, and the inner and outer spines are respectively indicated by the yellow and pink arrows in in \nfig{fig5}(c). In the magnetic pressure map, one can identify a small region of low magnetic pressure as indicated by the yellow arrow (\nfig{fig5}(e)), which represents the location of the null point structure. Measurement indicates that the height of the null point from the photosphere is about 9 Mm. Additionally, another region of high magnetic pressure below the east lobe of the fan manifests the location of the twisted mini-filament (see the pink arrow in \nfig{fig5}(e)).

\section{Conclusion and Discussion}
Using high resolution multi-wavelength observations taken by the NVST and the {\em SDO}, we studied a solar jet that was accompanied by the eruption of a mini-filament and a {\em GOES} C1.3 flare. The eruption led to an inner ribbon, an outer circular ribbon, and a remote ribbon. The fan-spine structure and the mini-filament underneath the fan are well revealed by the extrapolated three-dimensional coronal field, which indicate that the inner and remote ribbons were respectively located at the footprints of the inner and outer spines, while the circular ribbon was located at the footprint of the fan structure. During the rising phase of the flare, the inner ribbon and the east part of the circular ribbon showed a northward slipping motion. For the first time, we observed the interesting round-trip slipping motion of the west part of the circular ribbon, whose slipping directions were first toward to the X-shaped structure and then away from it along its original trajectory. The slipping motion of west part of the circular ribbon toward the X-shaped structure was at a constant speed of about \speed{898 $\pm$ 140}. The slipping motion away from the X-shaped structure showed a  deceleration of \accel{3.47 $\pm$ 0.55} and an initial speed of about \speed{663 $\pm$ 81}. In this event, the X-shaped structure observed in the H$\alpha$ images can be regarded as the projection of the coronal null point as evidenced in the extrapolated coronal field. It is also measured that the height of the coronal null points is about 9 Mm from the solar surface.

Analysis results indicate that the start time of the magnetic flux cancellation around the south end of the mini-filament was coincided with the beginning of the flare, which suggests that the initiation of the flare/jet eruption was probably caused by the flux cancellation, through breaking the connection of the mini-filament from the solar surface. Therefore, the eruption of the filament would be from south to north and squeezed the fan QSL and resulted in the slipping magnetic reconnection in the QSL, as well as the reconnection at the coronal null point. The accelerated particle beams that flowed down from the reconnection site along the reconnected magnetic field lines impact on the dense chromospheric and therefore formed the observed circular and inner bright ribbons. The northward slipping motions of the inner ribbon and the circular ribbon can be interpreted by the slipping magnetic reconnection in the fan QSL from south to north, while the slipping motion of the west part of the circular ribbon away from the X-shaped structure was due to the slipping reconnection in the fan QSL from north to the south. This can be expected during the recovery or oscillation stage of the magnetic system when the eruption of the filament weakened or stopped.

The jet appeared after the slipping motions of the flare ribbons, which had a speed of about \speed{100 (137)} in the H$\alpha$ (EUV) observations. During the eruption of the jet, brightenings at the X-shaped structure and the remote footprint of the outer spine were observed, in which the brightening at the X-shaped structure should be caused by the local heating of the plasma by the null point reconnection, while the remote flare ribbon at the footprint of the outer spine should be due to the impact of the particle beam accelerated by the null point reconnection along the outer spine. Regarding to the magnetic configuration of the fan-spine jet, it can be considered as the miniature version of the the large-scale breakout magnetic framework \citep[e.g.,][]{1999ApJ...510..485A,2012ApJ...750...12S,2017Natur.544..452W,2018ApJ...854..155K}. The significant similarities between small-scale fan-spine jets and large-scale breakout eruptions may hint at a scale-invariant of eruptive solar phenomenon.

Slipping motion of flare ribbons were not only observed in circular flare ribbons \citep[e.g.,][]{2009ApJ...700..559M,2013ApJ...774...60L,2015PASJ...67...78J,2017ApJ...837..173R,2017ApJ...850..167L,2018ApJ...859..122L}, but also in many other kinds of flare ribbons \citep[e.g.,][]{2014ApJ...784..144D,2016ApJ...823...41D,2016ApJ...830..152L,2016ApJ...823..136Z,2019ApJ...881...68L}. The slipping speeds reported in previous studies are ranged from \speed{30 to 450}. However, in our case the northward slipping speed of the circular ribbon's west part reached up to \speed{898 $\pm$ 140}, much higher than those observed in previous studies. The high slipping speed was probably due to the violent filament eruption underneath the fan structure, which squeezed the fan QSL and drove slipping magnetic reconnection within it. The slipping speed of the southward slipping motion of the circular ribbon's west part had slowed down to \speed{663 $\pm$ 81} and with a significant deceleration. This suggests that the erupting filament had weakened or stopped, and the magnetic system started a recovery or oscillation stage. Additionally,  most previous studies suggested that the slipping motions of flare ribbons are along one specific direction. \cite{2007Sci...318.1588A} reported the first simultaneous bi-directional slipping motions of flare ribbons caused by slipping magnetic reconnection in coronal loops. For circular flares, \cite{2012ApJ...760..101W} reported a case in which the inner ribbon showed a round-trip slipping motion. Slipping motions of flare ribbons are important for diagnosing three-dimensional magnetic reconnection. However, why and how the change of reconnection directions are still open questions. Further more observational and theoretical studies are desirable in the future.

\acknowledgments
The authors thank the excellent data provided by the NVST and {\sl SDO} teams, the anonymous referee's valuable comments and suggestions for improving the quality of the paper, and the helpful discussions with Dr. H. Li, and Dr. J. Hong from Yunnan Observatories. This work was supported by the Natural Science Foundation of China (11922307,11773068,11633008), the Yunnan Science Foundation (2017FB006), and the West Light Foundation of Chinese Academy of Sciences.

\end{document}